# Spontaneous Surface Collapse and Reconstruction in Antiferromagnetic Topological Insulator MnBi$_2$Te$_4$

Fuchen Hou[1]*, Qiushi Yao[1]*, Chun-Sheng Zhou[1], Xiao-Ming Ma[1], Mengjiao Han[1], Yu-Jie Hao[1], Xuefeng Wu[1], Yu Zhang[1,3], Hongyi Sun[1], Chang Liu[1,2], Yue Zhao[1,2]†, Qihang Liu[1,2,4]†, Junhao Lin[1]†

[1]Department of Physics, Southern University of Science and Technology, Shenzhen 518055, Guangdong, P. R. China
[2]Shenzhen Institute for Quantum Science and Engineering, Southern University of Science and Technology, Shenzhen 518055, P. R. China
[3]Department of Physics, University of Hong Kong, Hong Kong, P. R. China
[4]Guangdong Provincial Key Laboratory for Computational Science and Material Design, Southern University of Science and Technology, Shenzhen 518055, China

*These authors contributed equally to this work.
†Corresponding author. E-mail: linjh@sustech.edu.cn; liuqh@sustech.edu.cn; zhaoy@sustech.edu.cn

**Abstract**

MnBi$_2$Te$_4$ is an antiferromagnetic topological insulator which stimulates intense interests due to the exotic quantum phenomena and promising device applications. Surface structure is a determinant factor to understand the novel magnetic and topological behavior of MnBi$_2$Te$_4$, yet its precise atomic structure remains elusive. Here, we discovered a spontaneous surface collapse and reconstruction in few-layer MnBi$_2$Te$_4$ exfoliated under delicate protection. Instead of the ideal septuple-layer structure in the bulk, the collapsed surface is shown to reconstruct as Mn-doped Bi$_2$Te$_3$ quintuple-layer and Mn$_x$Bi$_y$Te double-layer with a clear van der Waals gap in between. Combining with first-principles calculations, such spontaneous surface collapse is attributed to the abundant intrinsic Mn-Bi antisite defects and tellurium vacancy in the exfoliated surface, which is further supported by in-situ annealing and electron irradiation experiments. Our results shed light on the understanding of the intricate surface-bulk correspondence of MnBi$_2$Te$_4,$ and provide insightful perspective of the surface-related quantum measurements in MnBi$_2$Te$_4$ few-layer devices.

**Introduction**

Magnetic topological quantum materials have stimulated intense research interest due to the interplay between magnetism and topology which results in emerging quantum phenomenon[1–4]. Examples include quantum anomalous Hall effect (QAHE)[3–5], Weyl semimetallic states[6,7], topological axion states[8] and Majorana fermions[2,9], etc, enabling potential applications in dissipationless electronic and quantum computing[10]. In the early research, a magnetic topological insulator (TI) is achieved by magnetically doping a TI thin film, in order to study the QAHE[3–5]. However, the random distribution of magnetic dopants introduces impurity scattering together with the ferromagnetic ordering, limiting the temperature for the realization of QAHE. Very recently, the tetradymite-type MnBi$_2$Te$_4$ compound was discovered as an intrinsic antiferromagnetic (AFM) TI in *A*-type AFM ground state with out-of-plane magnetic moments[11–19]. Although the existence of long-range magnetic order explicitly breaks the time-reversal symmetry, which is nevertheless preserved in conventional Z$_2$ TI[1,2], a new type of Z$_2$ invariant can be defined in MnBi$_2$Te$_4$, as long as a combined symmetry between time-reversal and fractional translation is preserved[11,20,21]. As a result, MnBi$_2$Te$_4$ provide an ideal platform of magnetic TI to realize QAHE and axion insulator state[15,17,22], etc.

On the other hand, there are still discrepancies between theoretical expectations and experimental facts in MnBi$_2$Te$_4$. For instance, theoretical predictions and some experimental observations declared a sizable magnetic gap at the surface of bulk MnBi$_2$Te$_4$[11–15], while recent report shows an unambiguously gapless Dirac cone at the (00$l$) surface of MnBi$_2$Te$_4$ crystal by using the high-resolution angle resolved photoemission spectroscopy (ARPES)[16,23,24]. Such inconsistency implies that the surface structure is a key factor requiring precise measurements, which may affect many of the corresponding novel magnetic and topological behavior in MnBi$_2$Te$_4$, such as the QAHE in odd-layers and the zero Hall plateau as an indicator of axion state in even-layers[11,17,18]. Apparently, most of the previous results consider the surface structure using the ideal septuple-layer (SL) MnBi$_2$Te$_4$ lattice [11–19], which, on the other hand, lacks direct proof to connect the bridge between theory and experiments.

Since the surface-bulk correspondence is the kernel of topological properties, in this paper, we systematically studied the atomic structure of the surface in MnBi$_2$Te$_4$ few layers with intended surface protection. Using cross-sectional scanning transmission electron microscopy (STEM) imaging and atomic electron energy loss spectroscopy (EELS), we unambiguously determine the surface of few-layer MnBi$_2$Te$_4$ to be Mn-doped Bi$_2$Te$_3$ quintuple-layer (QL) decorated with crystalline/amorphous Mn$_x$Bi$_y$Te double-layer (DL) rather than the ideal SL layered structure, *i.e.*, a spontaneous surface collapse and reconstruction occurred during the mechanical exfoliation. Such surface collapse is highly reproducible in all samples we measured. We further discovered that Bi-Mn anti-site defects were omnipresent in bulk $MnBi_2Te_4$. Combining the density functional theory (DFT) calculations, we unveiled the origin of the surface collapse in few-layer $MnBi_2Te_4$ as a result of the synergistic interaction between the Bi-Mn anti-site defects and surface tellurium vacancies that unavoidably formed even in inert gas environment due to trace of oxygen, which makes the as-observed reconstructed surface preferential in energy landscape. Similar surface collapse and reconstruction is reproduced by heating the intact SL surface exfoliated in ultrahigh vacuum as probed by atomic scanning tunneling microscopy (STM), and by electron irradiation in STEM which simultaneously monitor the dynamical surface collapse and reconstruction process atom-by-atom. These experiments further verified that the spontaneous surface collapse and reconstruction occurred in a well-controlled inert gas environment, is indeed induced by surface Te vacancy. To date, most of the few-layer $MnBi_2Te_4$ devices, in which the exotic quantum phenomena are observed, are prepared in inert gas environment, where we believe the as-observed spontaneous surface collapse can still be triggered by the trace of oxygen. Our findings of the spontaneous surface collapse and reconstruction not only set notes on the fabrication of $MnBi_2Te_4$ few-layer devices, but also bring new insights in understanding the emerging quantum phenomena in this intrinsic magnetic insulator.

**Results and Discussion:**

**Atomic structure of the collapsed surface in exfoliated $MnBi_2Te_4$ few layer**

MnBi$_2$Te$_4$ single crystal is a member of van der Waals (VDW) layered materials. A single VDW layer of MnBi$_2$Te$_4$ consists of alternating Bi-Mn-Bi layer intercalated by Te, forming a SL structure. Its antiferromagnetism originates from the middle Mn cations with a super-exchanged interlayer interaction between adjacent Mn layers. Figure 1a shows the XRD measurement of the parent MnBi$_2$Te$_4$ crystal with sharp and intense peaks that follow the (00$l$), $l$ = 3n, diffraction rule, which agree quantitatively with those in the standard PDF files. In addition, X-ray photoelectron spectroscopy (XPS) characterization on the MnBi$_2$Te$_4$ single crystal reveals the core level peaks of Mn 2p, Bi 4f and Te 3d (Fig. S1) without any impurity signal from other crystalline phases and elements, indicating the high quality of the parent single crystal. The fresh cleaved surface of MnBi$_2$Te$_4$ crystal was initially investigated by high resolution atomic force microscopy (AFM) operated in the same inert gas environment. A rough surface was seen with steps ranging from 2-6 Å (1-3 atomic layers) (Fig. S2), suggesting a possible disordered surface structure. We then performed cross-section study in order to obtain atomic information of the surface. The surface was intentionally protected by graphite, ensuring minimum surface degradation during cross-section sample fabrication (See methods).

Figure 1b shows a low-magnified high-angle annular dark field (HAADF) STEM image of the as-prepared MnBi$_2$Te$_4$ few-layer cross-section, taken from the [110] direction. The graphite appears dark contrast due to its relative light atomic weight in the STEM image. Zoom-in images with atomic resolution of the surface and bulk are exhibited in Figs. 1c and 1e, respectively. Comparison of the two readily shows an apparent deviation of the surface structure: it is a five-atom layers instead of seven in surface, similar to the atomic structure of TI $Bi_2Te_3$ (Te-Bi-Te-Bi-Te) viewed along the *c*-axis, known as QL structure (atomic model is shown in Fig. 1d). In contrast, the atomic structure of the bulk $MnBi_2Te_4$ (Fig. 1e) is consistent with the previous reports[11,12,14], showing a SL structure (Te1-Bi-Te2-Mn-Te2-Bi-Te1) as depicted by the atomic model in Fig. 1f. Quantitative energy dispersive spectrum (EDS) mapping (see methods) further revealed the chemical composition of the surface to be mainly consisted of Bi and Te with trivial Mn. The ratio of Mn, Bi and Te at the surface QL layer, highlighted by the red dashed rectangle in Fig. 1g, is estimated as ~ 0.3:1.8:3 (Fig. 1h), while reaching the normal 1:2:4 ratio below the surface (Fig. S3).

Above the QL structure, an amorphous layer with brighter contrast than the nearby graphite protection layer is observed. This amorphous layer separated from the QL surface with a clear VDW gap, as highlighted by red dashed lines in Fig. 1c. Moreover, crystalline structure with a DL height was occasionally observed to be embedded in such amorphous layer, which are highlighted by the white arrows. A larger view of the surface (Fig. 1g) further manifested that these tiny crystalline structures are omnipresent above the surface, yet gapped by amorphous layers in between. EDS results of such amorphous layer and embedded crystalline quantum island (highlighted by the red dashed rectangle in Fig. 1g) clearly reveal that, in contrast to the QL layer, the major element is Mn and Te while with trivial Bi. The element ratio among Mn, Bi and Te is ~ 0.7:0.2:1 (Fig. 1i).

The cross-sectional results pointed to a key finding that the surface of the as-prepared MnBi$_2$Te$_4$ few-layer sample underwent a spontaneous collapse even with intentional surface protection during the mechanical exfoliation in inert gas environment. Instead of the ideal SL model with chemical composition Mn:Bi:Te of 1:2:4, which is the case below the surface as verified by quantitative EDS, the realistic surface split into a Bi-rich QL plus a Mn-rich DL crystalline/amorphous layer, with a complementary elemental distribution as 0.3:1.8:3 and 0.7:0.2:1, respectively. Such surface collapse and reconstruction are highly reproducible in all samples we measured, even in a mild transfer without heating during the drop-down process of the protection graphite layer. Mn-doped QL Bi$_2$Te$_3$ is stable as reported previously[13], however, on the other hand the DL Bi-rich MnTe is theoretically unstable (See Supplementary Section II for more detail), thus turning into amorphous phase once formed. The existence of the island-like DL crystalline structure is presumably due to the VDW interaction from the underlying layers and the local chemical composition fluctuation during the surface reconstruction.

**Detection of the Mn-Bi exchange antisite defects by atomic EELS and intensity quantification analysis**

The unexpected surface collapse and reconstruction may be related to the superficial chemical stoichiometry variation or surface defects[25]. To further find out the hint of the surface collapse, we took the atom-by-atom electron energy-loss spectroscopy (EELS) across the reconstructed surface. Figure 2a shows EELS of each atomic column in the surface QL structure, with the simultaneously collected HADDF image shown on right. The identity of Bi shown by the $M_4$ edge at 2688 eV only shows up at the $2^{th}$ and $4^{th}$ layer, corresponding to the two brightest spots, which is consistent to the expected STEM HAADF intensity due to its large atomic number. The atomic EELS and simulated STEM HAADF image (see Fig. S5) suggested the surface QL is alternating Te-Bi-Te-Bi-Te QL similar to the $Bi_2Te_3$ structure. However, it is surprising that clear Mn signal, labelled by the $L_{2,3}$ edge at 640 eV, also shows up exactly in the Bi columns. This suggests that most of the Mn atoms doped into the Bi lattice column given the high miscibility at the cation sites, i.e., antisite defects, $Bi_{Mn}$ or $Mn_{Bi}$, may present in the surface even before the surface collapse occurred.

To further confirm the scenario of $Bi_{Mn}$ or $Mn_{Bi}$, we also collected atomic EELS from the bulk SL structure. Figure 2b shows the EELS across the SL structure. Surprisingly, we found clear Bi signal appeared in the Mn layer ($4^{th}$ atomic column in Fig. 2b), while strong Mn signal at both the Bi layers ($2^{nd}$ and $6^{th}$ layer), evidencing the presence of $Bi_{Mn}$ and $Mn_{Bi}$. As a result, the Mn layer has much brighter intensity than the simulated STEM HAADF image which used the ideal SL model (Fig. S5), due to the inclusion of the heavier Bi atoms, and vice versa. In contrast, Te layers show no Mn or Bi signals, excluding the presence of other types of anti-site defects. However, elemental distribution of a SL in the bulk is exactly 1:2:4 as probed by EDS, which implied the occurrence of intralayer exchange between Bi and Mn atoms, resulting in almost equal amount of $Bi_{Mn}$ and $Mn_{Bi}$ in both Mn and Bi layers. Since such Mn-Bi intralayer exchange is omnipresent in the bulk, it played a key role in the surface collapse and reconstruction process.

In order to further investigate the Mn-Bi intralayer exchange effect, we mapped the intensity of all atomic columns in SL structure in a large scale using a peak intensity finding software[26] and performed quantitative statistical analysis. Figure 2c shows the intensity histogram. Mn, Bi and Te sites are mapped separately, as indicated by the markers shown on the representative STEM image. To quantitatively study the intralayer exchange between the Mn and Bi layers, we compared the experimental value to the simulations. The dark grey dashed lines in Fig. 2c are the simulated intensity of Mn, Bi and Te columns using the ideal SL structure without any anti-site defects, where the Te column is normalized to the experimental value for direct comparison. As expected, intensity of all Mn column is much higher than the simulated one while Bi is lower, both of which had wider distribution than Te, a direct evidence of intralayer Mn-Bi intermixing. From the simulated intensity, when the concentration of $Bi_{Mn}$ reaches 30% in the Mn layer, the intensity of the Mn atomic column is almost similar to that of Te (see Fig. S5c). Therefore, the average concentration of the anti-site defects in both Mn and Bi layers can be qualitatively estimated by the full-with-half-maximum (FWHM) of the intensity distribution with comparison to the simulation with different concentration of Mn-Bi intermixing. The $Bi_{Mn}$ concentration in Mn-layer is found to be 40%~50%, which is approximately twice value of $Mn_{Bi}$ concentration in Bi atomic layers. This is consistent with our previous analysis that the amount of $Bi_{Mn}$ and $Mn_{Bi}$ should be equal due to the intralayer exchange, while Bi is double the amount of Mn due to the 1:2:4 chemical stoichiometry.

**Physical origin of the defect-triggered surface collapse and reconstruction**

To uncover the physical origin of the as-observed surface reconstruction in MnBi$_2$Te$_4$, we presented comprehensive thermodynamic defect calculations by using density functional theory (DFT). By far, our experiments evidenced that high concentration of intrinsic anti-site defects (Bi$_{Mn}$ and Mn$_{Bi}$) are presented in the Mn and Bi layers, thus we firstly examine the formation of anti-site defects due to the Bi-Mn exchange in MnBi$_2$Te$_4$. Chemical potential substantially affects the calculations of defect formation energy, therefore we determine the accessible range of the chemical potential, *i.e.*, growth condition, of MnBi$_2$Te$_4$ in ($\Delta\mu_{Mn}$, $\Delta\mu_{Bi}$) parameter space with the constraints imposed by competing binary compounds, as shown in the green area of Fig. 3a. In the unstable regions (white area), MnBi$_2$Te$_4$ tends to decompose to various competing phases. Therefore, the formation energies of native defects are calculated merely under two representative environments, i.e., Bi-rich condition ($\Delta\mu_{Bi} \approx 0$ eV), and Te-rich condition ($\Delta\mu_{Te} \approx 0$ eV) denoted by A and B points in Fig. 3a, respectively. More information about defect calculations is provided in Supplementary Section III to IV.

For Bi-rich condition (see Fig. 3b), Bi$_{Mn}$, having the lowest formation energy, is the dominant donor defect due to the excess valence electrons of Bi than Mn. On the other hand, the cation-to-cation antisite defect (Mn$_{Bi}$) has much lower formation energy than the anion-to-cation antisite defect (Te$_{Bi}$) even in Te-rich condition (Fig. S7), thus Bi$_{Mn}$ and Mn$_{Bi}$ are two dominant defects in MnBi$_2$Te$_4$. The combination of these two antisite defects creates a double defect of Mn-Bi exchange (Mn$^{Bi}$), with relatively small formation energies shown in the dark-green line of Fig. 3b and 3c. This explained the high concentration of the cation-to-cation antisite exchange defects observed in the STEM cross-section image. Note that to form $Mn^{Bi}$, no atoms exchange between $MnBi_2Te_4$ and reservoirs is required.

The physical origin of ideal SL to the as-observed reconstructed surface (QL+DL structure) should be closely related to the defect landscape of $MnBi_2Te_4$. Firstly, we consider the total energy of $MnBi_2Te_4$ with $Mn^{Bi}$ exchange defects at the surface, and compare with the total energy of $Bi_2Te_3$ and MnTe islands under surface reconstruction. The latter is set to 0 as the reference, denoted by the dashed line in Fig. 3e. We find that the energy of the defective SL surface increases monotonically with the increasing $Mn^{Bi}$ concentration, which is consistent with its positive defect formation energy from our calculation (see supplementary Section V for calculation models and methods). To pass the reference line and thus realize the collapse from SL to QL, the required concentration of $Mn^{Bi}$ is extremely high (over 50%). Such a high defect concentration is not observed in our experiments because the bulk still has stable SL framework against reconstruction, only surface did. In other words, although $Mn^{Bi}$ is the dominant defect under equilibrium growth condition, $Mn^{Bi}$ alone can hardly promote the SL surface collapsed into QL. Therefore, the driving force of the surface collapse should be something else, most likely some defects which may form at the surface regardless of their relatively large formation energies in bulk.

From bulk to surface, the outermost layer (Te1 layer in $MnBi_2Te_4$) suffers the strongest environmental perturbation, such as unintentionally introduced atomic vacancies upon cleavage. It is known that tellurides like $MnBi_2Te_4$ is readily to be oxidized, which would break the equilibrium growth condition. Taking tellurium oxides into account, we reevaluate the formation energy of Te1 vacancy $V_{Te1}$ (see supplementary Section VI for calculation details). As illustrated in Fig. 3d, under equilibrium Te-poor condition, the formation energy of $V_{Te1}$ is about 1.1 eV, indicating the relatively low $V_{Te1}$ concentration in bulk. However, with the assistance of oxygen, instead of forming elemental Te solid, Te forms tellurium oxides with $V_{Te1}$ formation energy decreased dramatically. We consider three tellurium oxides, $TeO_2$, $Te_2O_5$ and $TeO_3$. For all cases, the calculated $V_{Te1}$ formation energies are about −2 eV. The total energy of $MnBi_2Te_4$ with $V_{Te1}$ defects at the surface increases quickly with the increasing of $V_{Te1}$ concentration, as shown in Fig. 3e. To realize SL to QL collapse, the required surface $V_{Te1}$ concentration is about 20% in Bi-rich condition. With the assistance of $Mn^{Bi}$, the surface collapse may take place at an even lower $V_{Te1}$ concentration. Overall, as illustrated in Fig. 3f, oxygen at the surface reacts with Te1 sublayer, leaving tellurium vacancies. Such surface $V_{Te1}$ serves as the dominant driving force to trigger the surface collapse and reconstruction in $MnBi_2Te_4$ by accelerating Bi-Mn exchange. As a result, the residual Te and Mn atoms form MnTe islands covering QL surface sparsely. This is indeed the case that we observed small amount of oxygen signal at the interface between the QL structure and the graphite, as detected both by EELS and EDS (see Fig. S4 and Fig. 1i), which confirmed the contribution of oxygen in creating surface Te vacancies by forming oxides of tellurium. Moreover, the theory also suggests that such surface reconstruction involve a kinetic thermal unequilibrium process which would result in an incomplete Bi-Mn exchange, consistent of the chemical stoichiometry of the QL and the amorphous/crystalline DL structure as probed by EDS (See Fig1h and 1i).

**Verifying the defect-induced surface collapse and reconstruction by in-situ STM and STEM**

To verify the scenario of the Te-vacancy-driven surface collapse suggested by theory, we first tried to exclude the presence of oxygen by investigating in-situ cleaved $MnBi_2Te_4$ surface using ultrahigh vacuum STM (operation pressure better than $2 \times 10^{-10}$ Torr), in which the oxygen concentration is multiple orders of magnitude lower than the inert gas environment. Figure 4a shows the STM image of a freshly exfoliated $MnBi_2Te_4$ single crystal terminating with the (00*l*) surface. Abrupt steps with uniform height of ~1.4 nm (7 atomic layers, Fig. 4a) was obtained. Zoom-in STM image shows an atomic crystalline surface (Fig.4b and 4c) with randomly distributed dark spots. According to the bias-voltage-dependent STM images of these dark spots (see Fig. S9), they are ascribed to the $Bi^{Mn}$ antisite defects underneath the Te layer, consistent with our STEM results and the previous report[27]. This indeed confirmed that the ideal SL $MnBi_2Te_4$ surface can be preserved in ultrahigh vacuum.

To introduce Te vacancies at the surface, an in-situ heating at 150 °C was then applied to the surface for about 41 hours, since Te is easily sublimated at elevated temperature. A clear structure collapse is observed, evidenced by the emergence of holes with step height about 0.4 nm (2 atomic layers, Fig. 4d) inside the originally flat surface (see Fig. 4e). As a result, the height of the collapsed region, as shown by the dimmer contrast in Fig. 4e, is about 1 nm (the height profile in Fig. 4d), corresponding to five atomic layers. The collapsed surface still shows similar atomic crystalline structure with hexagonal patterns (Fig. 4h and 4g) due to the surface similarity between $MnBi_2Te_4$ and $Bi_2Te_3$.

The above results are in agreement with the surface collapse and reconstruction mechanism suggested by theory. We also attempted to directly capture the dynamical process of the collapsed surface reconstruction. It is well-known that high energy electron being used in imaging also simultaneously transfers momentum and energy to the specimen which can result in beam-induced defects[28]. Figures 5a-5c show sequential STEM images as a function of electron dose, highlighting the dynamical surface collapse and reconstruction process in $MnBi_2Te_4$ surface. As the electron dose accumulated, the atomic column intensity of the outermost layer (Te1 layer in QL structure) is dimmed due to the loss of tellurium (Fig. 5b). Meanwhile, the gap between the surface QL and the next SL, which is underneath the Te loss region as highlighted by the arrow in Fig. 5b, narrowed down from 2.6 Å to 2 Å. A significant split is seen in the first two atomic layers of the SL, a trend in separation into a DL and QL structure. Finally, as the two outermost atomic columns are eliminated by electron bombardment, the surface structure collapsed and reconstructed from a QL+SL to double QLs (Fig. 5c).

The dynamical collapse and reconstruction process indicate that the VDW gap can close and reopen during the formation of defects depending on the exact layer configurations. This indeed means that the VDW gap between layers can be discontinued in a reconstructed surface as long as the outmost surface is a QL structure, which is highlighted by the white arrows in Fig. 5d. Moreover, once Te vacancy was formed, the ideal SL could not be the surface layer but only the QL did, due to the Te-vacancy induced rapid exchange between the Mn/Bi layer, leaving a reconstructed DL and QL as the outmost surface. This is evidenced in Fig. 5e, which shows a large view of collapsed and reconstructed surface including QL+ SL, two QLs and DL + QL structures, all of which manifested QL as the stable surface structures. The in-situ dynamical imaging unambiguously confirmed that the formation of Te vacancy lead to the surface collapse and reconstruction in $MnBi_2Te_4$ single crystal.

**Discussion**

Our results reveal that the surface structure of $MnBi_2Te_4$ is not as stable as previously thought, a condition that inevitably affects the surface electronic structure and thus the topological surface-bulk correspondence. Theoretically, when the surface inherits the crystal and magnetic structures of the bulk, a gapped Dirac cone with dozens of meV is expected due to the intrinsic magnetism. However, high-resolution ARPES measurements unambiguously show a robust gapless Dirac cone at $MnBi_2Te_4$ surface[16,23,24]. One possibility is that the local moments of Mn atoms tends to distribute randomly, giving rise to the almost zero band gap with linear dispersion. However, direct experimental evidence, such as surface magnetic configurations, is still lacking to support this hypothesis. Here, we suggest from our experimental results that the surface collapse leads to significant absence of the magnetic atoms together with magnetic disorder, resulting in the vanishing surface gap. First of all, with the surface MnBi$_2$Te$_4$ SL collapsing to Bi$_2$Te$_3$ QL, the origin of the Dirac gap opening is mainly the proximity effect from the ordered magnetic moments of the second topmost SL. Compared with the perfect SL surface, the proximity-induced gap is much smaller. To confirm this, we calculated by DFT the surface electronic structure with both terminations, i.e., perfect MnBi$_2$Te$_4$ SL and collapsed Bi$_2$Te$_3$ QL. As shown in Fig. S10, compared with the MnBi$_2$Te$_4$ termination without surface collapse and reconstruction, the surface band gap of MnBi$_2$Te$_4$ with the surface layers degraded to Bi$_2$Te$_3$ shows a significant reduction (from 42 meV to 5 meV). In addition, residual MnTe DL islands on Bi$_2$Te$_3$ QL surface tend to be antiferromagnetic with parallel Mn spins in the basal plane[29,30]. Therefore, the floating Mn atoms in the DL atoms cannot help to open the surface Dirac gap either. Finally, our findings suggest that the device application of thin-film MnBi$_2$Te$_4$, e.g., the quantum anomalous Hall effect, may also suffer the impacts of surface collapse and reconstruction, which calls for further exploration.

**Conclusion**

In summary, we have discovered that a spontaneous surface collapse and reconstruction in exfoliated MnBi$_2$Te$_4$ single-crystal occurs even under the protection of a well-controlled inert gas environment. Combing STEM imaging, STM experiments and DFT calculations, we systematically show such surface collapse resulted from the synergistic effect of the high-concentrate intrinsic Mn-Bi exchange defects and the formation of tellurium vacancy on the surface, which is induced by the trace of oxygen in the inert gas environment. The surface reconstruction and the existence of massive intrinsic defects bring a more comprehensive understanding of the antiferromagnetism and the anomalous quantum states of $MnBi_2Te_4$ few-layer devices. The sensitive surface also set a note on all $MnBi_2Te_4$ few-layer device fabricated in non-ultrahigh vacuum environment, shed light in understanding the surface-related measurement of transport, and exploration of exotic quantum phenomena and device fabrication for applications based on $MnBi_2Te_4$ crystal.

**Materials and Methods**

**1. Sample fabrication**

The few-layer $MnBi_2Te_4$ was exfoliated from bulk $MnBi_2Te_4$ through a scotch-tap method in a glove box filled with argon. The parent $MnBi_2Te_4$ bulk crystal is grown by flux method[16]. The fresh surface is exposed in argon atmosphere, and subsequently covered by graphite through a routine dry transfer method in the glove box to encapsulate the surface from being oxidized. The cross-section STEM specimens were quickly prepared using Focused Ion Beam after the sample was fetched from the glove box, all of which ensured minimum surface degradation.

**2. Characterizations**

**XRD -** Single-crystal X-ray diffraction was performed on a Rigaku Miniex diffractometer using Cu K$\alpha$ radiation at room temperature.

**XPS -** The X-ray Photoelectron Spectroscopy measurement on the freshly exfoliated surface of $MnBi_2Te_4$ crystal was performed on PHI 5000 Versaprobe III. The spectrum was analyzed by the PHI-MultiPak software.

**AFM -** Atomic Force Microscopy (AFM) measurement was carried out using the Asylum Research, Cypher S system placed in an inert gas environment. To minimize the oxidation of $MnBi_2Te_4$, the exfoliation of the $MnBi_2Te_4$ crystal and AFM measurements are performed one after another in the same glove box.

**STM -** The STM experiments were carried out with a low-temperature STM (UNISOKU Co., Ltd., USM1500) in ultrahigh vacuum (UHV) condition. The $MnBi_2Te_4$ single crystal was cleaved along the (00*l*) crystal plane in the STM chamber with a base pressure of $2\times10^{-10}$ mbar. The freshly cleaved sample was immediately transferred to the STM chamber for further measurements at 78 K (or 5K). The sample was baked at 150 ºC for 41 hours to introduce the surface reconstruction. The tungsten tip was prepared by electrochemical etching and subsequent ebeam heating and Ar+ sputtering. We trained the tip apex on clean Cu (111) surface prior to all measurements. STM topography images were processed by WSxM[31].

**STEM –** STEM imaging, EDS and EELS analysis on $MnBi_2Te_4$ crystal were performed on a FEI Titan Themis with a X-FEG electron gun and a DCOR aberration corrector operating at 60 kV. The inner and outer collection angles for the STEM images ($\beta_1$ and $\beta_2$) were 48 and 200 mrad, respectively. The convergence semi-angle of the probe is 25 mrad. The beam current was about 100 pA for high angle annular dark-field imaging, the EDS and EELS chemical analyses. All imaging was performed at room temperature. The quantitative element ratio of $MnBi_2Te_4$ crystal was confirmed by inductively coupled plasma mass spectrometry (ICP-MS) analysis. The ratios between Mn, Bi and Te were normalized based on the bulk $MnBi_2Te_4$ crystal. Thereinto, in the electronic irradiation experiment, the electron dose (*D*) is calculated by *D* = *I* × *T*/*A*, using the beam current (*I*), beam illuminating area (*A*), and the irradiation time (*T*).

## 3. Density Functional Theory Calculations

First-principles calculations were carried out using Vienna *ab initio* simulation package (VASP)[32] within the framework of density functional theory (DFT)[33]. Exchange-correlation functional was described by the generalized gradient approximation with the Perdew-Burke-Ernzerhof (PBE) formalism[34]. The electron-ion interaction was treat by projector-augmented-wave (PAW) potentials[35] with a planewave-basis cuff of 500 eV. The whole Brillouin-zone was sampled by Monkhorst-Pack grid[36] for all models. Due to the correlation effects of 3*d* electrons in Mn atoms, we employed GGA+U approach within the Dudarev scheme and set the U to be 5 eV, which was investigated by and previous work[16]. All atoms were fully relaxed until the force on each atom was less than 0.01 eV/Å and the total energy minimization was performed with a tolerance of $10^{-5}$ eV. Freely available software VASPKIT[37] was used to deal with VASP output files. The calculation process is detailed in Supplementary Section II to VI.

**Supporting Information**

Supporting Information for this article is available in the online version of the paper.

Section S1. Characterization of the exfoliated surface of $MnBi_2Te_4$ crystal

Section S2. Stability of pure and Bi-doped MnTe double-layer

Section S3. Thermodynamic limits on the chemical potentials in DFT calculation

Section S4. Defect formation energy calculations

Section S5. Relative surface energy calculations

**Acknowledgements**

We thank J. Zhang for the support of XPS measurement. The authors would like to acknowledge the support from National Natural Science Foundation of China (Grant No.11974156, 11874195 and 11674150), Guangdong International Science Collaboration Project (Grant No. 2019A050510001 and 2017ZT07C062), National key research and development program (Grant No. 2019YFA0704901), the Guangdong Provincial Key Laboratory of Computational Science and Material Design (Grant No. 2019B030301001), the Key-Area Research and Development Program of Guangdong Province (2019B010931001), Guangdong Innovative and Entrepreneurial Research Team Program (Grant No. 2016ZT06D348) and also the assistance of SUSTech Core Research Facilities, especially technical support from Pico-Centre that receives support from Presidential fund and Development and Reform Commission of Shenzhen Municipality. First-principles calculations were also supported by Center for Computational Science and Engineering at SUSTech.

J.L. conceived the project. F.H. and J.L. made the TEM samples, performed AFM measurement and STEM related experiments, analysis and simulations. DFT calculations were done by Q.Y., H.S. and Q.L. Sample growth and X-ray analysis was made by X.M., Y.H. and C.L. STM measurement was carried out by C.Z., X.W., Yu Z. and Y.Z. J.L. constructed the schematic of MnBi$_2$Te$_4$ surface collapse and reconstruction. M.H. participated in parts of STEM experiments. The work was coordinated by J.L., Q.L. and Y.Z. The manuscript was written by J.L., F.H., Q.L. and Q.Y. with input from all authors. All authors commented on the manuscript.

**Figures**

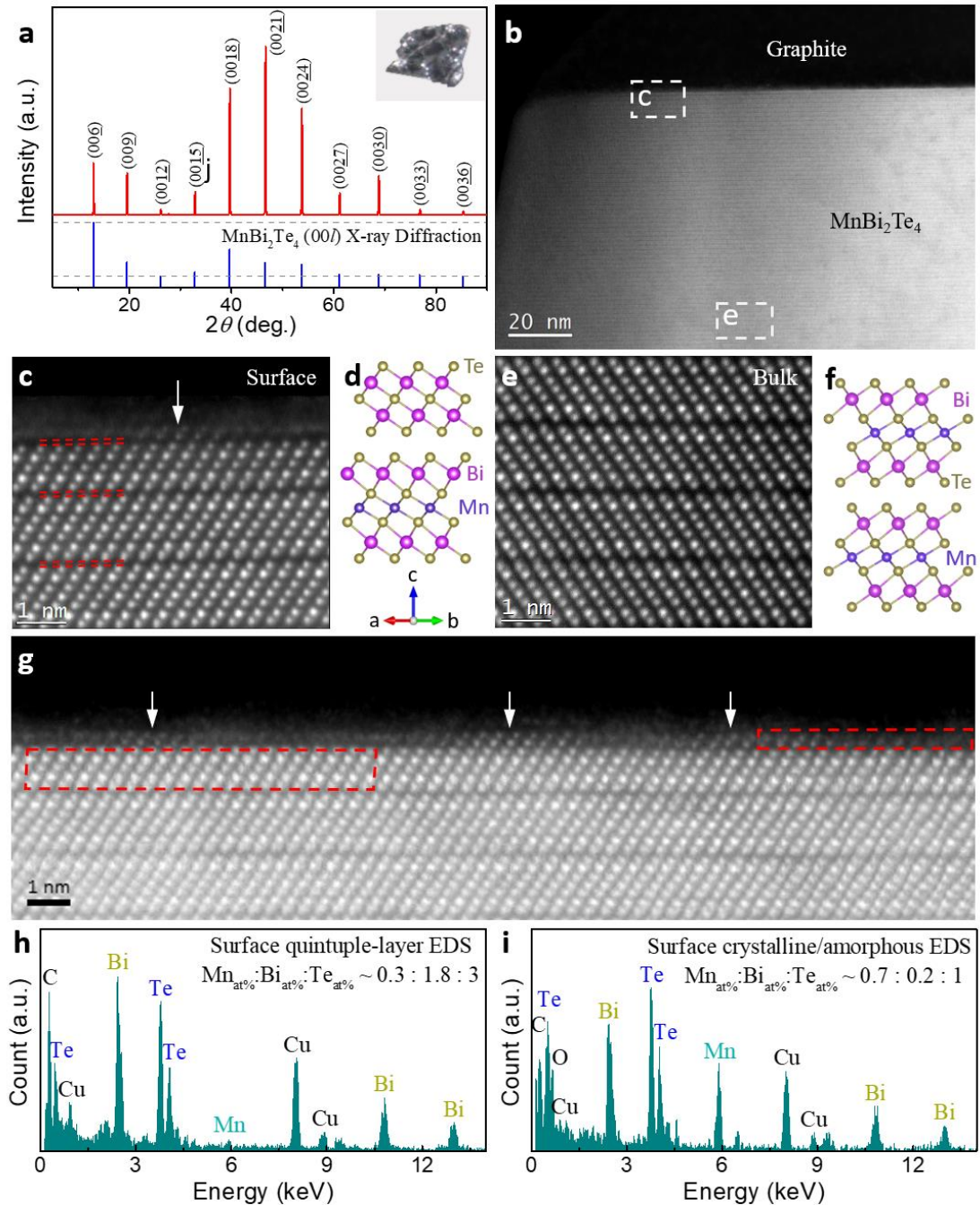

**Figure 1 Atomic characterizations of the MnBi$_2$Te$_4$ surface structure.** (a) X-ray diffraction (XRD) pattern (red), and the referenced (00*l*) peaks (blue) from standard PDF file of the parent single MnBi$_2$Te$_4$ crystal. (b) Large-scale high-angle annular dark field scanning transmission electron microscopy (HAADF-STEM) image of the cross section of layered MnBi$_2$Te$_4$ crystal viewed along the [110] direction, with the overlaid graphite as surface protection. (c-e) Zoom-in atomic resolution HAADF-STEM images of the surface and bulk highlighted in (b), respectively. A quintuple-layer (QL) with double-layer (DL) crystalline/amorphous structure, instead of the ideal septuple-layer (SL) MnBi$_2$Te$_4$, is seen at the surface. The arrows highlight the crystalline DL islands on the surface. (d) and (f) are the corresponding atomic models from the images, respectively. (g) A large scale of the atomic structure of the MnBi$_2$Te$_4$ surface showing the omnipresent crystalline DL islands indicated by arrows. (h, i) The corresponding energy dispersive spectrum (EDS) maps for the surface QL and DL crystalline/amorphous structure, respectively. Cu and C come from the grid and substrate.

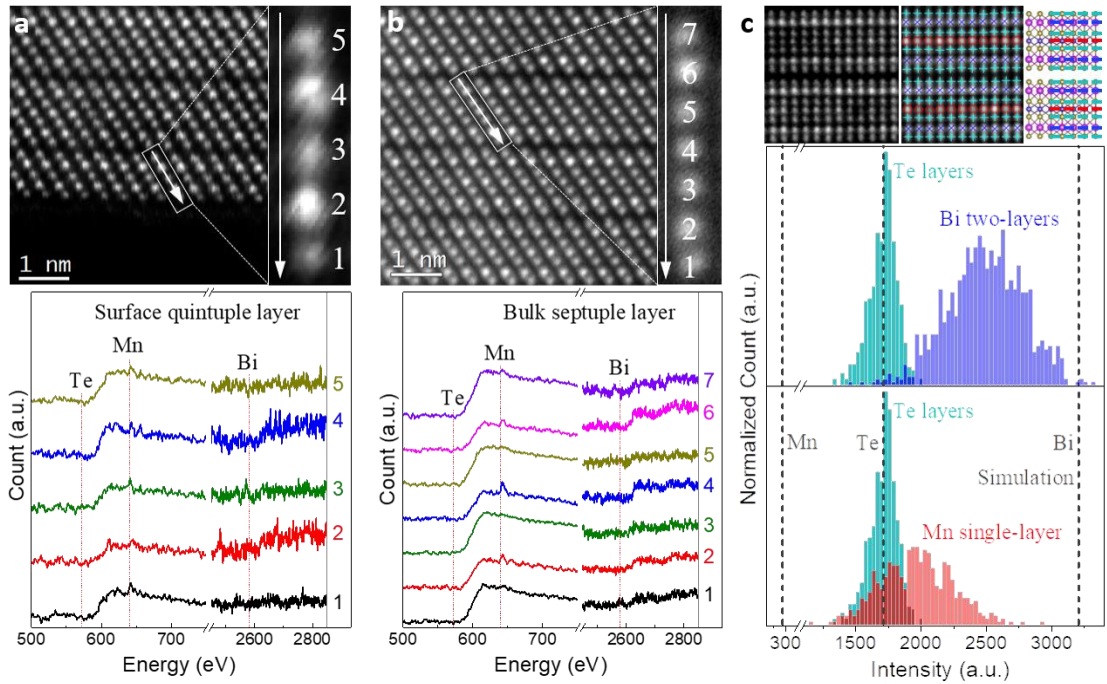

**Figure 2. Chemical analysis of the exchange Mn-Bi defects at the surface and bulk.** (a-b) Atomic resolution cross-sectional HAADF images of single-crystal MnBi$_2$Te$_4$ with the arrow indicating the position and direction of the electron energy loss spectrometry (EELS) linescan acquisition. The corresponding background subtracted atom-by-atom EELS data for the surface QL (a) and bulk SL (b) are shown below. The numbers mark the different atomic columns corresponding to the labeled EELS data. The onset energy of Te, Mn and Bi are set as guide to eye. (c) Histogram of the intensity distribution mapped from the Te (cyanine), Bi (blue) and Mn (red) atomic columns in bulk MnBi$_2$Te$_4$, respectively. The HAADF image shows the location of different atomic columns marked by corresponding colors, view from the $[1\bar{1}0]$ direction. The dark grey dashed lines represent the normalized intensity of Te, Mn and Bi columns from simulation without any Mn-Bi exchange defect.

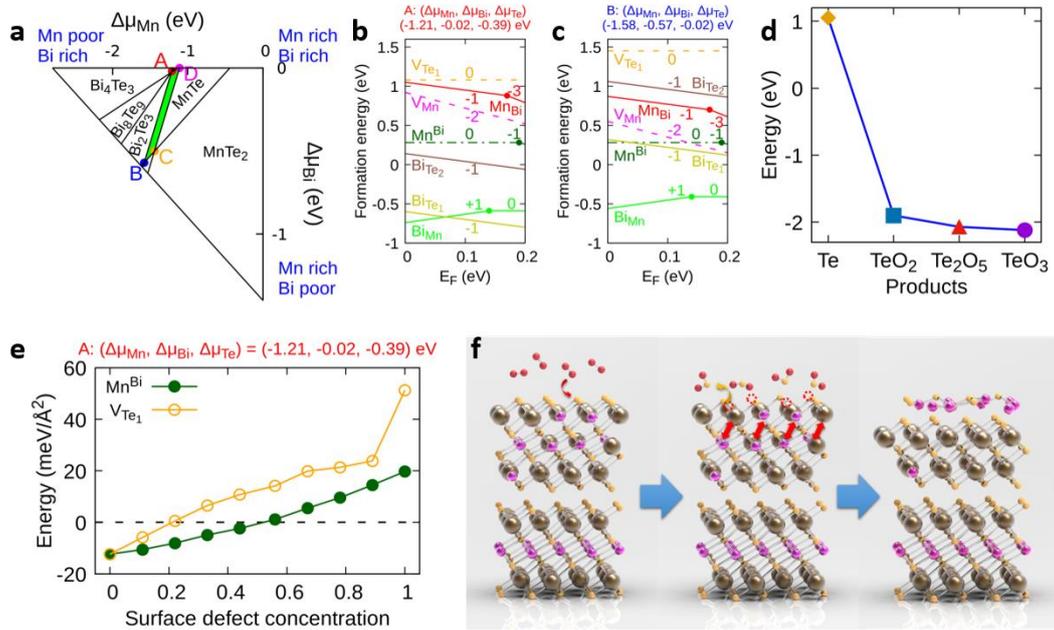

**Figure 3. Physical origin of the surface collapse in MnBi$_2$Te$_4$.** (a) The allowed chemical potential domain (green area) for MnBi$_2$Te$_4$ shown in ($\Delta\mu$Mn, $\Delta\mu$Bi) parameter space, which is sketched out by points A, B, C and D. The other regions are excluded due to the formation of competing phases specified in the figure. (b, c) Formation energies of defects in MnBi$_2$Te$_4$ for chemical potential sets A and B shown in (a). Formation energy at C and D are shown in the Supplementary Information. (d) Calculated formation energy of V$_{Te1}$ for different tellurium oxides. (e) Relative surface energy as a function of defect concentration under chemical potential set A. Dashed line indicates the energy of surface terminating with Bi$_2$Te$_3$ and MnTe islands. (f) Schematic of the surface collapse and reconstruction in MnBi$_2$Te$_4$ crystal induced by the formation of oxygen-driven Te vacancy and subsequent Mn-Bi exchange effect.

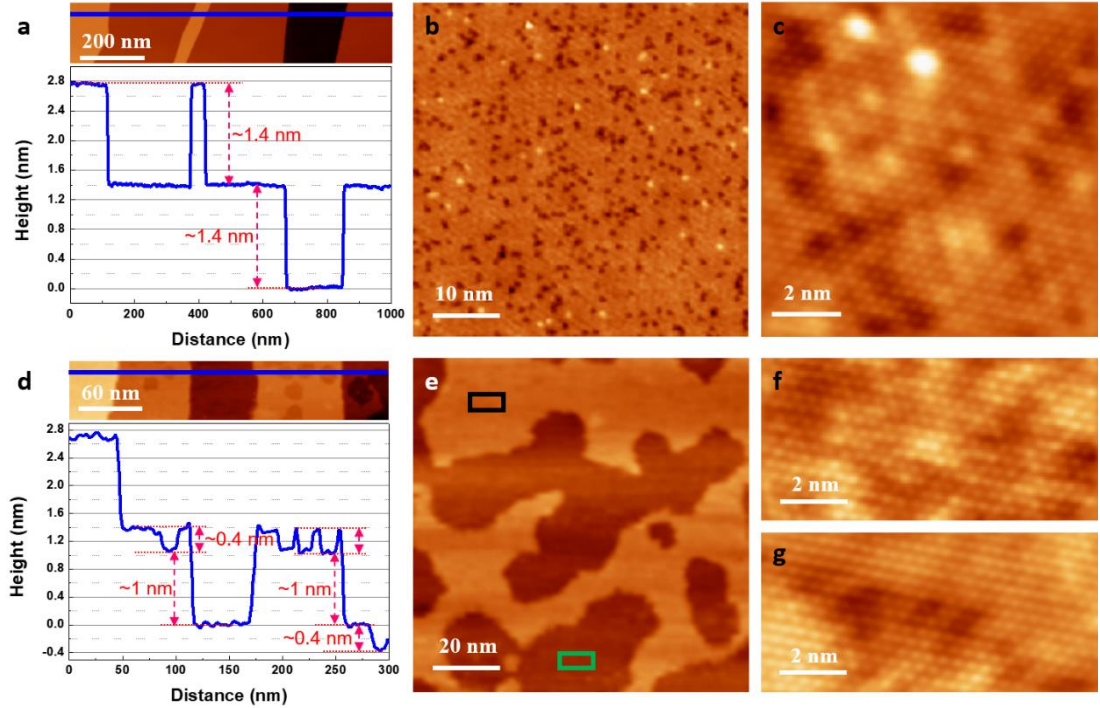

**Figure 4. In-situ heating of the exfoliated MnBi$_2$Te$_4$ surface in ultrahigh vacuum.** (a) Large-scale STM image of MnBi$_2$Te$_4$ (00$l$) surface exfoliated in ultrahigh vacuum and the corresponding height line profile along the blue line. (b) Zoom-in STM image of the MnBi$_2$Te$_4$ (00$l$) surface with defects. (c) Atomic-resolution image of the MnBi$_2$Te$_4$ surface. The dark spots are concluded as the Bi$^{Mn}$ anti-site defects under the Te layer. The SL height and the crystalline surface imply the intact MnBi$_2$Te$_4$ surface can be preserved in ultrahigh vacuum. (d) Large-scale STM image of the in-situ heated MnBi$_2$Te$_4$ surface and the corresponding height line profile along the blue line. Holes with DL heights appear after heating. (e) Zoom-in image of the annealed surface. (f, g) Atomically resolved STM images of the annealed surface marked by black and green box in (e), respectively. Tunneling Parameters: (a) V$_{bias}$ = 1V, I$_t$ = 200 pA (b) V$_{bias}$ = 1V, I$_t$ = 50 pA (c) V$_{bias}$ = -1 V, I$_t$ = 3 nA (d) V$_{bias}$ = 1.2 V, I$_t$ = 20 pA (e) V$_{bias}$ = 1.5 V, I$_t$ = 1 nA (f) V$_{bias}$ = -1 V, I$_t$ = 1 nA (g) V$_{bias}$ = 0.2 V, I$_t$ = 500 pA.

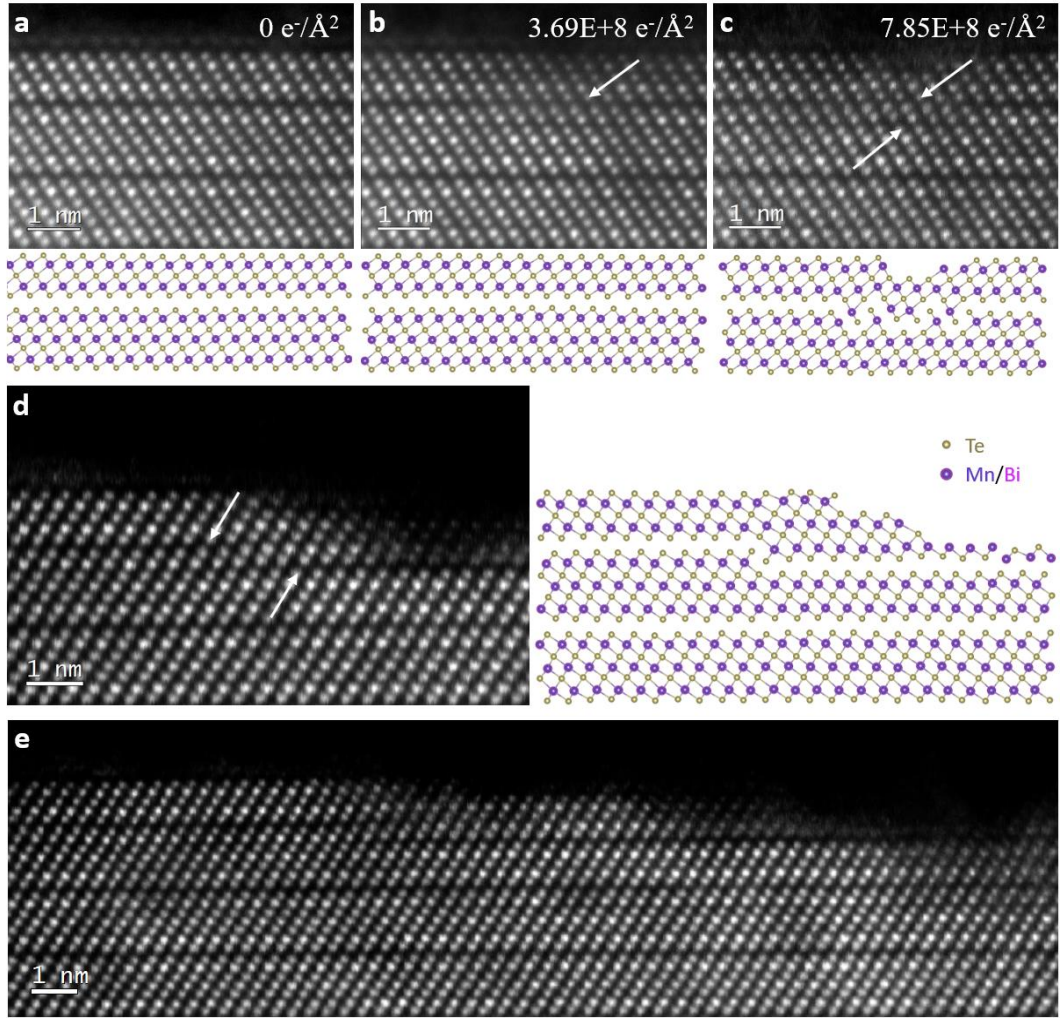

**Figure 5. In-situ observation of the dynamical surface collapse and reconstruction process induced by electron irradiation.** (a-c) Evolution of the surface atomic structure as a function of electron dose: 0 e$^-$/Å$^2$ (a), 3.69E+8 e$^-$/Å$^2$ (b) and 7.85E+8 e$^-$/Å$^2$ (c). The corresponding atomic models are shown below where the Bi and Mn atoms are not discriminated due to the massive existence of Mn$^{Bi}$ exchange defects. The arrows in (c) indicate the close of the original VDW gap due to the loss of Te atoms, with a new VDW gap opening two layers down the collapsed and reconstructed region. (d) Atomic STEM image showing a discontinue VDW gap along the surface after collapse and reconstruction. (e) STEM image showing a large scale of surface structure after prolonged electron irradiation, where all the surfaces were terminated by the QL or QL+DL structures. Noted that the collapse and reconstruction process occurred under ultra-high vacuum.

# Supporting Information

## Section I. Characterization of the exfoliated surface of MnBi$_2$Te$_4$ crystal

To identify the elemental composition of the parent MnBi$_2$Te$_4$ crystal, X-ray photoelectron spectroscopy (XPS) measurement was performed on the fresh surface which was exfoliated in the high purity argon protection in the glove box. The XPS results displays that the parent crystal consists of Mn, Bi and Te without any impurity elements (Fig. S1), indicating the high quality MnBi$_2$Te$_4$ single crystal. In addition, the low O 1s peak demonstrates surface oxidation of MnBi$_2$Te$_4$. This can be further certified by the corresponding high-resolution (HR) spectra of individual Mn, Bi and Te elements. As the HR-spectra showed, the Mn 2p peak could be reconstructed into multi-sub peaks, which arises from different bonding including Mn-Te and Mn-O bonds according to the binding energy of different peaks. For Bi spectrum, two peaks are ascribed to Bi 4f$_{7/2}$ and Bi 4f$_{5/2}$ due to Bi-Te bonds. In Te 3d spectrum, two major peaks at 571.7 eV (Te 3d$_{5/2}$) and 582.1 eV (Te 3d$_{3/2}$) represent Bi-Te bonds and Mn-Te bonds, respectively. Thereinto, two minor peaks at 575.6 eV and 586 eV can be observed in Te 3d spectrum, which is attributed to Te-O bonds. These results indicate that the surface structure is highly sensitive which should not be exposed to ambient atmosphere (*1*).

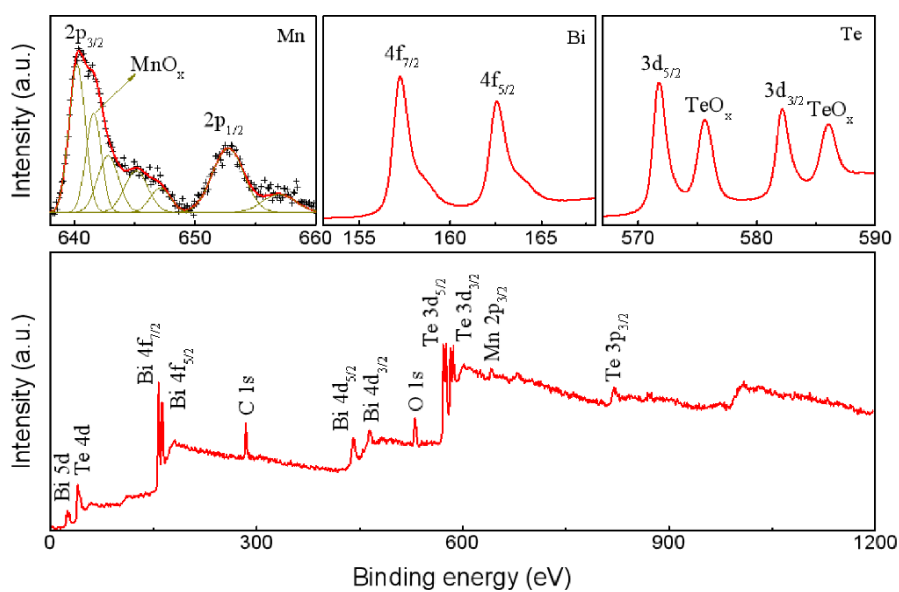

**Supplementary Figure 1 Chemical analysis of MnBi$_2$Te$_4$ single crystal at the cleaved surface.** Survey X-ray photoelectron spectroscopy (XPS) spectra of fresh surface of MnBi$_2$Te$_4$ (bottom) and corresponding high-resolution Mn 2p, Bi 4f and Te 3d spectra (top).

To observe the initial surface structure, the high-resolution atomic force microscopy (AFM) was performed to study the freshly exfoliated surface of MnBi$_2$Te$_4$ crystal under the inert gas protection in the glove box. Figure S2 shows the AFM image measured by contact mode, indicating a rough surface. The height linescan acquisition is highlighted by the arrow, indicating the position and direction. We can clearly observe that the height of steps ranging from 2-6 Å (as shown in the height curve), implying a possible highly disordered surface structure.

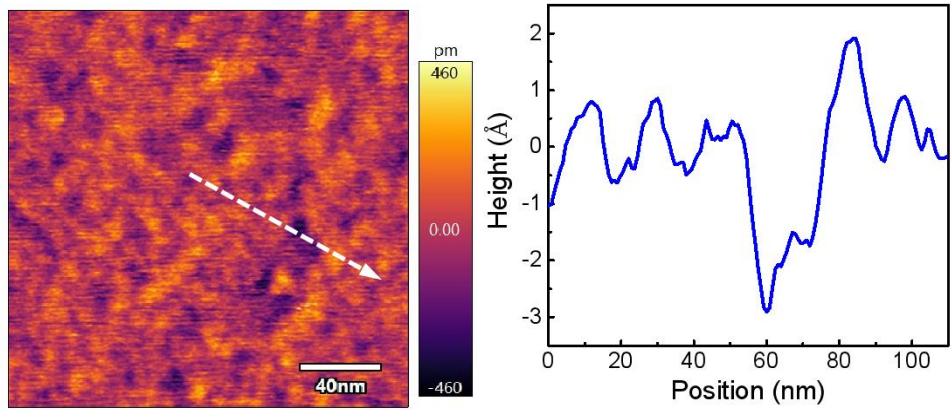

**Supplementary Figure 2 Characterization of the cleaved surface morphology.** Atomic Force Microscopy (AFM) image on the fresh cleaved surface of MnBi$_2$Te$_4$ crystal with the arrow indicating the position and direction of the height linescan acquisition.

To quantitative identify the chemical composition, we studied the energy dispersive spectrum (EDS) of the bulk MnBi$_2$Te$_4$ crystal, which below the surface, highlighted by the red dashed rectangle in Fig. S3a. The quantitative EDS mapping verifies the bulk is MnBi$_2$Te$_4$ single crystal, with estimated ratio of Mn, Bi and Te as ~ 1:2:4 in the septuple-layer structure (see Fig. S3b). In addition, the EDS elemental maps for individual elements of Mn, Bi and Te indicate that manganese mainly distributes in the middle of septuple layer in the corresponding STEM image, bismuth and tellurium elements mainly distribute on the corresponding stacking layers in bulk SLs as shown in Fig. S3c.

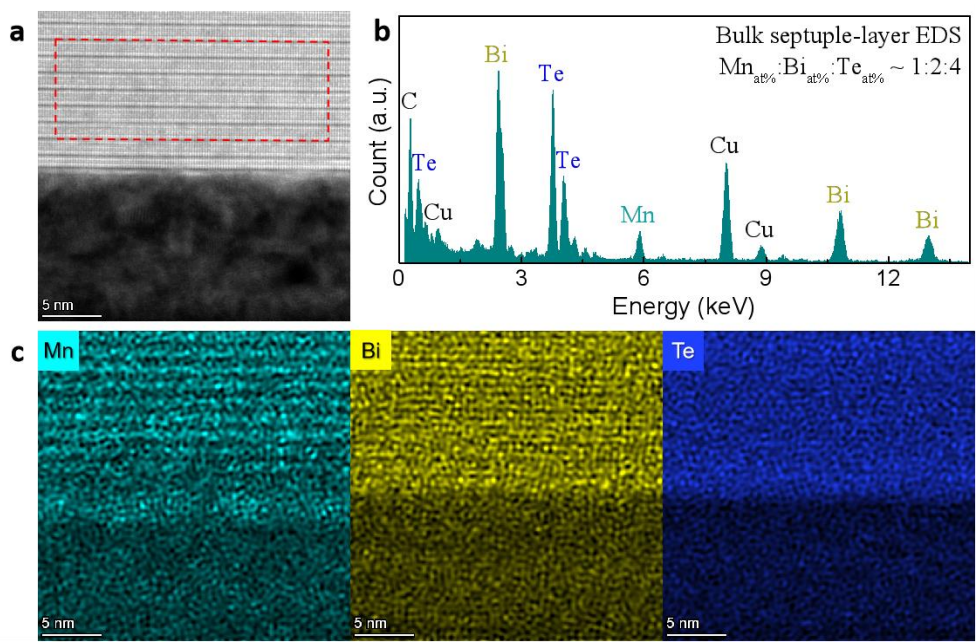

**Supplementary Figure 3 Chemical analysis of MnBi$_2$Te$_4$ single crystal at surface and bulk.** (a) HAADF image of atomic structure of single MnBi$_2$Te$_4$ at cross section of surface, taken from the [1$\bar{1}$0] direction. (b) The energy dispersive spectrum (EDS) from the highlighted zone by the red dashed rectangle in (a). Cu and C come from the grid and substrate, respectively. (c) The corresponding EDS maps for individual elements of Mn, Bi and Te.

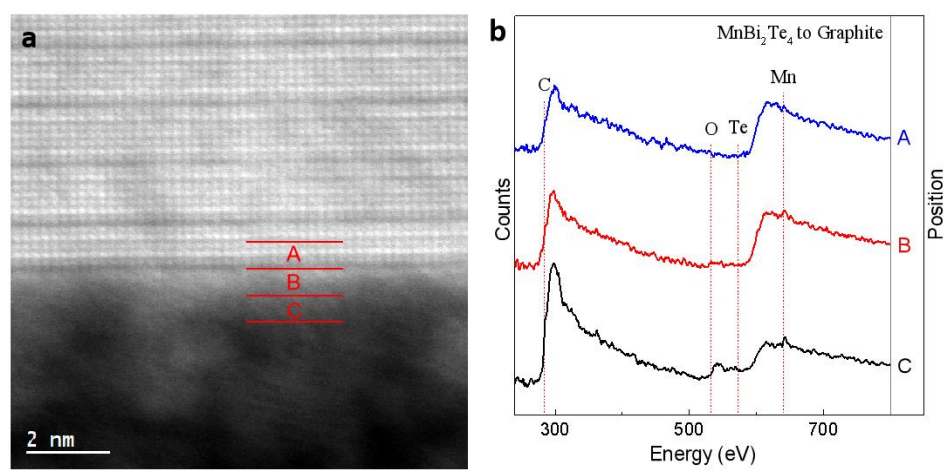

**Supplementary Figure 4 Chemical analysis of the cleaved surface with the protection using graphite.** (a) HAADF image of atomic structure of single $MnBi_2Te_4$, (b) corresponding EELS maps for individual area from (a), confirming the position of $MnBi_2Te_4$ and Graphite.

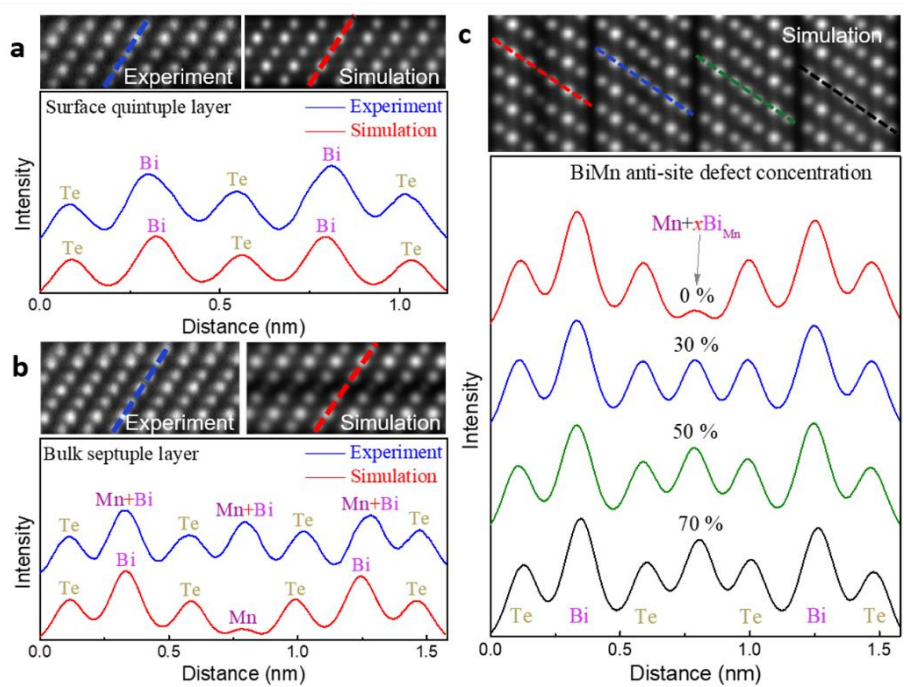

**Supplementary Figure 5 The intensity variation in STEM image with different concentration of exchange Mn-Bi defects.** (a-b) Line intensity profiles along the highlighted blue (experiment) and red (simulation using the ideal models) dashed lines in the corresponding images, indicating the intensity of each atomic columns in the surface quintuple-layer (a) and bulk septuple-layer (b). (c) Line intensity profiles along the $MnBi_2Te_4$ septuple-layer with different Bi-Mn anti-site defect concentrations. Red (0%), blue (30%), green (50%) and dark (70%) dashed lines highlighted the intensity of each atomic columns with different Bi-Mn exchange defects in the simulation.

In order to verify the composition of the double crystalline/amorphous layer

(called DL-region) on the exfoliated surface of MnBi$_2$Te$_4$ crystal, we collected the EELS from different region of the surface, including QL region, DL-region and the region above DL-region as shown in Fig. S4. Obviously, the graphite is above the DL-region according to the abrupt enhancement of C signal in *C*-area compared to *A*- and *B*-area. The QL has slight Mn with almost invisible Mn signal in *A*-area, which is consistent with the EDS results (Fig. 1h). In the *B*-area, we found clear Mn, Te and some O signals. Slight Mn and Te signal are observed in graphite region presumably due to the sputtering during FIB sample preparation.

The atom-by-atom EELS confirmed the existence of Mn$^{Bi}$ anti-site exchange defects in bulk SL structure. To quantify the defects concentration, simulation was carried out using models with different concentration of Mn$^{Bi}$ anti-site exchange defects. Figure S5 shows the intensity of each atomic column along the SL structure as highlighted dashed lines. We found discrepancy between experiment and simulation in the bulk MnBi$_2$Te$_4$ crystal, especially for the Mn atomic layer in the SL structure as shown in Fig. S5b. The simulation results display a very low *Z*-contrast of Mn layer, however the intensity is higher than the Te layer in experiment. Combined with the EELS results in Fig. 2b, it can be proved that the middle layer of SL has a large number of Bi$_{Mn}$ anti-site defects. To quantify the concentration of anti-site defect in experiment, we performed the simulation on the SL MnBi$_2$Te$_4$ single crystal with a series of Bi$_{Mn}$ anti-site defects concentration. The simulation results display that the intensity of Mn layer is equal to the Te layer as the Bi$_{Mn}$ anti-site defects concentration reaches 30% (see Fig. S5c). Therefore, we can estimate concentration of anti-site defects in MnBi$_2$Te$_4$ crystal based on their intensity.

**Section II. Stability of pure and Bi-doped MnTe double-layer**

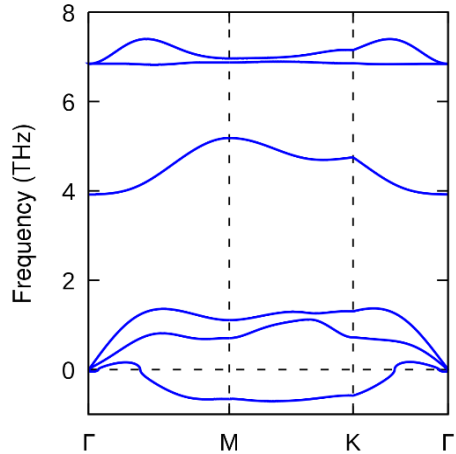

**Supplementary Figure 6 The calculations on the stability of MnTe DL.** Phonon dispersion of MnTe DL.

Phonon dispersion calculations were carried out to study the stability of pure and Bi-doped MnTe double-layer (DL). Here, open source package Phonopy(*2*) was used for phonon calculations with forces obtained from Vienna *ab initio* simulation package (VASP)(3) by finite displacement method. As shown in Fig. S6, phonon dispersion of MnTe DL contains sizable imaginary eigenfrequencies through symmetry lines, which indicates that MnTe DL is dynamically unstable. On the other hand, Bi-doped MnTe DL (Bi:Mn:Te = 0.22:0.78:1) underwent dramatically structure deformation during optimization. Such significant reconstruction indicates that the supposed Bi-doped MnTe in DL framework is energetically rather unstable.

**Section III. Thermodynamic limits on the chemical potentials in DFT calculation**

We determine the allowed chemical region with restrictions below, followed by the approach proposed in previous reports(4, 5).
(i) To maintain a stable MnBi$_2$Te$_4$ compound, the sum of chemical potentials of its constituent atoms must equal the formation enthalpy of the compound. That is
$$\Delta\mu_{Mn} + 2\Delta\mu_{Bi} + 4\Delta\mu_{Te} = \Delta H(MnBi_2Te_4) \qquad (1)$$
(ii) In addition to eq. (1), to avoid solid/gas elemental precipitation, we need
$$\Delta\mu_{Mn} \leq 0, \Delta\mu_{Bi} \leq 0, \Delta\mu_{Te} \leq 0, \qquad (2)$$
Combination of eq. (1) and eq. (2) sketch out the entire chemical potential domain.
(iii) Constraints are also imposed by other possible competing phases. For example, to avoid forming Bi$_2$Te$_3$, we need
$$2\Delta\mu_{Bi} + 3\Delta\mu_{Te} \leq \Delta H(Bi_2Te_3) \qquad (3).$$
Here, we consider as competing phases Bi$_4$Te$_3$, Bi$_8$Te$_9$, Bi$_2$Te$_3$, MnTe and MnTe$_2$. The resulting accessible region of the chemical potentials is illustrated in Fig. 3a in ($\Delta\mu_{Mn}$, $\Delta\mu_{Bi}$) plane.

**Section IV. Defect formation energy calculations**

The formation energy of a defect $\alpha$ in charge state $q$ is defined as(6, 7)
$$H(\alpha^q) = E(\alpha^q) - E(host) - \sum_i n_i\mu_i + q(E_F + \varepsilon_{VBM}^{host}) + E_{corr.} \qquad (4)$$

Where E(host) is the energy of the pure host supercell, and E($\alpha^q$) is the total energy of defective structure using an equivalent supercell. The integer $n_i$ indicates the number of atoms of type $i$ (host atoms or impurity atoms) that have been added to ($n_i > 0$) or removed from ($n_i < 0$) the supercell to form the defect, and the $\mu_i$ are the corresponding chemical potentials of these species. E$_F$ ranges from 0 to bulk bandgap. $\varepsilon_{VBM}^{host}$ is the valence band maximum eigenvalue of the host. Finally, E$_{corr.}$ is a correction term account for finite-size effects(8–10). In this work, we calculate the correction energy with the freely available SXDEFECTALIGN code(10). Based on A, B, C and D sets of chemical potentials shown in Fig. 3a, we calculate typical native defect formation energies, as shown in Fig. S7. Note that, under all equilibrium growth conditions, Bi$_{Mn}$ remains the dominant defect with negative formation energy. In fact, all allowed growth conditions impose $\Delta\mu_{Mn} < -1$ eV, which means Mn rather poor condition. Mn sublattice is readily to be occupied by Bi or Te atoms. However, formation energy of Te$_{Mn}$ is larger than that of Bi$_{Mn}$ by more than 1 eV even in Te rich condition. Taking this into account, for all growth conditions, Bi$_{Mn}$ is the dominant defect in MnBi$_2$Te$_4$, which may not easy to be ruled out by manipulating growth condition.

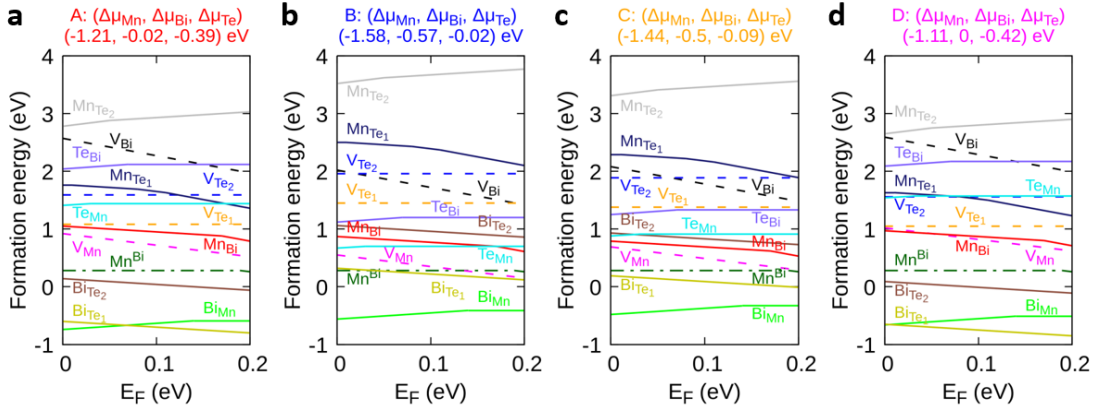

**Supplementary Figure 7 DFT calculations on the formation energies of various defects.** Formation energies of defects in MnBi$_2$Te$_4$ for chemical potential sets A, B, C and D shown in Fig. 3a.

**Section V. Relative surface energy calculations**

To simulate MnBi$_2$Te$_4$ surface, we construct slab model contains two Te1-Bi-Te2-Mn-Te2-Bi-Te1 septuple layer (SL). One SL is fixed while the other one is relaxed. To decouple the interaction between neighboring slabs, a 15 Å thick vacuum layer was added to the slab along the *c* direction. As the reference, we set energy of slab with surface terminating by Bi$_2$Te$_3$ and islands MnTe to 0 (dash line in Fig. S8). For slab with surface terminating by SL framework including defect, the energy is calculated by energy of slab with defective surface plus total chemical potentials of defect atoms. For example, considering surface including $xV_{Te1}$, its energy is defined as

$$E(slab_{T\text{-}SL}^{xV_{Te1}}) + x\mu_{Te1} \tag{5}$$

Where $E(slab_{T\text{-}SL}^{xV_{Te1}})$ is energy of slab terminating by SL containing $xV_{Te1}$.

The relative surface energy is then obtained by

$$E(slab_{T\text{-}SL}^{xV_{Te1}}) + x\mu_{Te1} - E(slab_{T-Bi_2Te_3}) - E(MnTe) \tag{6}$$

Where $E(slab_{T-Bi_2Te_3})$ is energy of slab terminating by Bi$_2$Te$_3$ layer. E(MnTe) is energy of bulk MnTe. The calculated relative surface energy for points A, B, C and D in Fig. 3a are shown in Fig. S8.

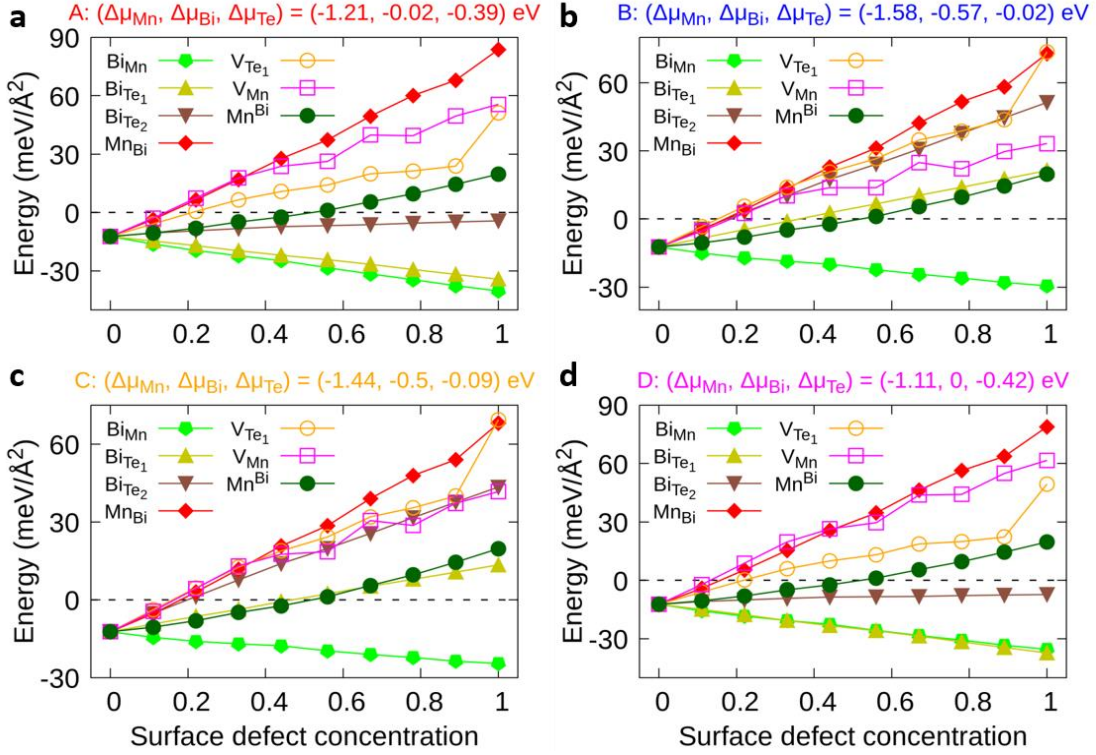

**Supplementary Figure 8 DFT calculations on relative surface energy.** Relative surface energy as a function of defect concentration under chemical potential set A, B, C and D shown in Fig. 3a.

**Section VI. Te1 vacancy formation energy with different final products**

For different final products, $V_{Te1}$ formation energies are calculated by

$$E(\text{defect}) + \frac{1}{m}\mu_{Te_mO_n} - E(\text{host}) - \frac{n}{2m}\mu_{O_2} \qquad (7)$$

Where E(defect) is energy of supercell including one $V_{Te1}$. E(host) is energy of perfect equivalent supercell. $\mu_{Te_mO_n}$ is chemical potential of tellurium oxide $Te_mO_n$. $\mu_{O_2}$ is chemical potential of oxygen.

**Section VII. STM images with different bias voltage**

From the STM images (Fig. 5), we can observe massive defects on the cleaved surface of MnBi$_2$Te$_4$ single crystal. A series of bias voltage was applied to examine the defects (see Fig. S9). The images were taken at the same location with identical tunneling current of 0.5 nA. As bias voltage changes from positive to negative, the dark spots remain their contrast, changing from triangularly located dots to blurred dark spot, similar to the previous reported Bi$^{Mn}$ anti-site defects in Mn doped Bi$_2$Te$_3$ crystal.

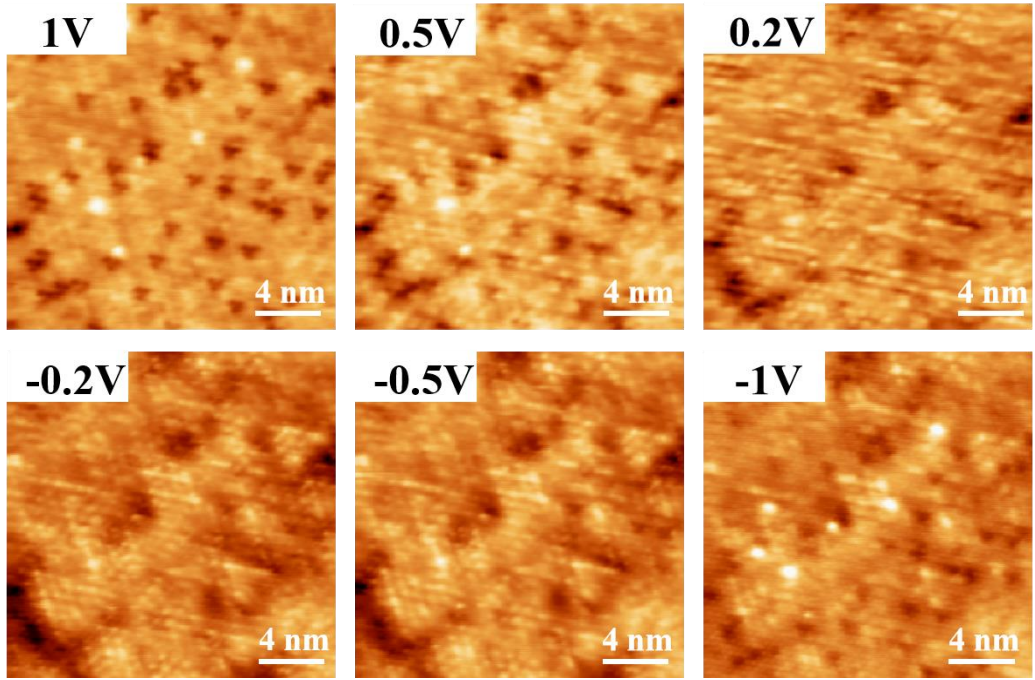

**Supplementary Figure 9 In-situ characterizations of the cleaved surface morphology.** Sample-bias-dependent STM topographs of the freshly cleaved MnBi$_2$Te$_4$ (00$l$) surface.

## Section VIII. Band structures of slabs with SL and QL terminations

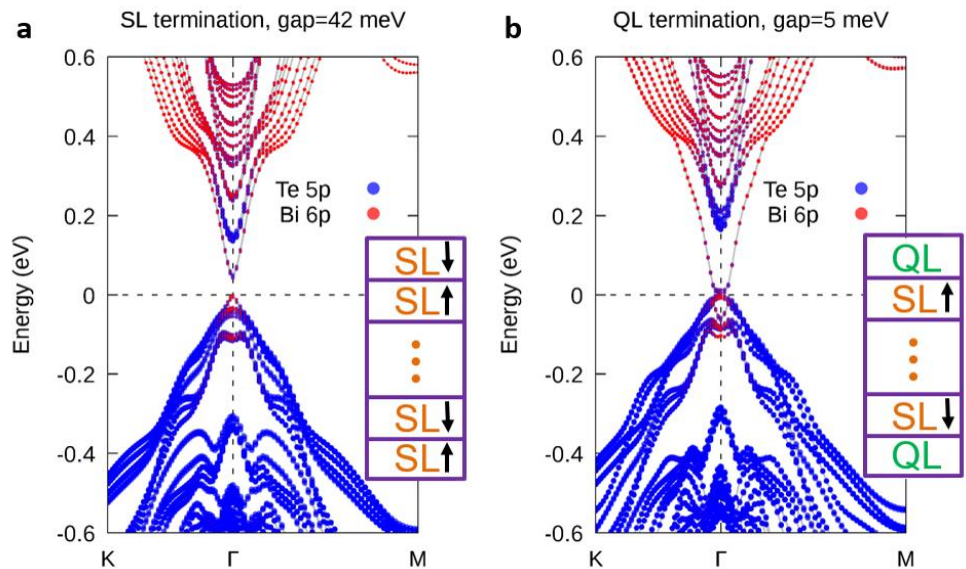

**Supplementary Figure 10 DFT calculated surface band structures of MnBi$_2$Te$_4$ SL and Bi$_2$Te$_3$ QL terminations.** a. Calculated band structure of an 8-SL slab. b. Calculated band structure of a slab composed of 6 SLs sandwiched by 2 QLs. Te 5p and Bi 6p states are indicated by blue and red points, respectively. Slabs models are schematized by the bottom right rectangles. An A-type AFM spin configuration is used in our simulations, as indicated by the black arrows in the rectangles.